\documentclass[aps,prl,twocolumn,floatfix]{revtex4}
\usepackage{amssymb}
\usepackage{amsmath}
\usepackage{graphicx}

\begin{document}

\title{Magnetic susceptibility of crystals with crossing of their  band-contact lines}

\author{G.~P.~Mikitik}
\affiliation{B.~Verkin Institute for Low Temperature Physics \&
Engineering, Ukrainian Academy of Sciences, Kharkov 61103,
Ukraine}

\author{Yu.~V.~Sharlai}
\affiliation{B.~Verkin Institute for Low Temperature Physics \&
Engineering, Ukrainian Academy of Sciences, Kharkov 61103,
Ukraine}

\begin{abstract}
We theoretically study the orbital magnetic susceptibility produced by electron states near a crossing point of two band-contact lines in a crystal. It is shown that this susceptibility can have an unusual dependence on the Fermi level and can change noticeably  with the temperature when the Fermi level is in the vicinity of the crossing point. These features of the magnetic susceptibility can be useful in detecting the crossing points in crystal. The obtained results explain the well-known temperature dependence of the magnetic susceptibility of V$_3$Si.
\end{abstract}

\maketitle

\newpage

\section{Introduction}

Band-contact lines, i.e., the lines along which two electron energy bands touch in a Brillouin zone, exist in majority of crystals with a weak spin-orbit interaction \cite{herring,m-sh14,kim,fang}. For example, such lines were found in Bernal graphite \cite{graphite}, beryllium \cite{beryl,beryl1}, magnesium \cite{beryl1}, aluminium \cite{al}, LaRhIn$_5$ \cite{prl04}, and in the bulk Rashba semiconductors BiTeI and BiTeCl (see, e.g., \cite{m-sh19}). The band-contact lines (nodal  lines) also exist in all the topological nodal-line semimetals which are intensively studied at present  \cite{m-sh19,armit,bernevig,gao,weng-r,fang-r}.
However, the band-contact lines can cross each other at some points in symmetry axes of crystals. The crossing of two band-contact lines, which can occur in twofold or fourfold symmetry axes, was theoretically studied recently \cite{m-sh20}. In particular, the crossing of this type takes place in Mackay-Terrones crystals \cite{weng}, ZrB$_2$ \cite{lou18,wang18}, V$_3$Si \cite{gork,step}, and in the ZrSiS-family of the nodal-line semimetals \cite{schoop,pez,fu19,chen17,hosen,pan,delft,guo19}.

Appearance of a new pocket of the Fermi surface or a break of its neck are the well-known $2\frac{1}{2}$-order electron topological transitions of Lifshitz \cite{lif,lak,var,var1}. In the Brillouin zone, in the vicinity of points of these transitions, the appropriate electron band disperses quadratically in the quasi-momentum ${\bf p}$. Appearance of self-intersecting Fermi surfaces with changing the chemical potential $\zeta$ is another type of the electron topological  transition. This transition takes place in crystals with the band-contact lines, and it is usually of the $3\frac{1}{2}$ kind \cite{m-sh14,m-sh-jltp}. This kind of the transition results from the fact that in the vicinity of an {\it isolated} band-contact line (in a plane perpendicular to it), the two contacting bands disperse linearly in the quasi-momentum ${\bf p}$ \cite{m-sh16,m-sh18} rather than quadratically. However, in the vicinity of the crossing point, the dispersion relation for the two contacting  bands essentially changes as compared to the case of the isolated band-contact line \cite{step,m-sh20} [see also Eqs.~(\ref{2}) and (\ref{3}) below]. This change leads to the fact that the appearance of the  self-intersecting Fermi surfaces near the crossing point is the topological transition different from the $3\frac{1}{2}$-order one \cite{step,m-sh20}. The characteristic feature  of this transition  is that the Fermi-surface transformation is accompanied by an unusual  dependence of the magnetic susceptibility on the chemical potential $\zeta$ \cite{step},
\begin{eqnarray}\label{1}
 \chi=\Delta \chi \left[1+\exp(\frac{\varepsilon_{\rm cr}-\zeta}{T})\right ]^{-1}+\chi_0,
 \end{eqnarray}
when the magnetic field $H$ is directed along one of the band-contact line at the crossing point. Here the constant factor $\Delta \chi$ depends on the parameters characterizing this point, $\varepsilon_{\rm cr}$ is energy of the crossing point, and $\chi_0$ is a constant background that is independent of $\zeta$ and of the temperature $T$. This $\chi_0$ is due to the electron states that are far away from the crossing point. It should be also noted that the first term in formula (\ref{1}) results from the orbital motion of electrons in the crystal.

In this paper we develop results of Ref.~\cite{step} and present a new explanation of the published experimental data on the magnetic susceptibility of V$_3$Si.

\section{Electron spectrum near the crossing point}

Neglecting the spin-orbit interaction, the electron energy spectrum for the two bands ``$c$'' and ``$v$'' in the vicinity of the crossing point of the two band-contact lines has the form \cite{step}:
 \begin{eqnarray}\label{2}
 \varepsilon_{c,v}({\bf p})\!\!&=&\!\varepsilon_{\rm cr}+a p_3+B_1 p_1^2 + B_2 p_2^2\pm E({\bf p}), \\
 E({\bf p})\!\!&=&\!\left[(a'p_3+B_1'p_1^2+B_2'p_2^2)^2+ \beta^2p_1^2p_2^2\right]^{1/2}\!\!\!, \label{3}
 \end{eqnarray}
where the $p_3$ axis in the quasi-momentum space coincides with the symmetry axis in which the crossing point occurs; the axes $p_1$ and $p_2$ are along the tangents to the band-contact lines at their crossing point;  all $p_i$ are measured from this point; $a$, $a'$, $B_i$, $B_i'$, $\beta$  are constant parameters, and $\varepsilon_{\rm cr}$ is the energy of the crossing point.
For the four-fold symmetry axis when $B_1=B_2$ and $B_1'=B_2'$, the $E({\bf p})$ can contain the term $\beta^2(p_1^2-p_2^2)^2$ instead of $\beta^2p_1^2p_2^2$. However, this case reduces to Eq.~(\ref{3}) by the rotation of the coordinate axes $p_1$, $p_2$ by the angle $\pi/4$. The band-contact lines are determined by the condition $E({\bf p})=0$, and Eq.~(\ref{3}) specifies these two lines as follows: $p_2=0$, $p_3=-B_1'p_1^2/a'$ and $p_1=0$, $p_3=-B_2'p_2^2/a'$, with the point $p_1=p_2=p_3=0$ being their crossing point.

The parameter $a$ describes the tilt of the spectrum (\ref{2}) along the  symmetry axis. At present, we are not aware of a noticeable  tilt of the spectrum along such an axis for any material, i.e., the parameter $a$ seems small for all the known crossing points. In this regard, we shall assume below that $a= 0$ in Eq.~(\ref{2}). This assumption simplifies the subsequent formulas without imposing fundamental restrictions on the results.
In the cubic crystal V$_3$Si, the crossing point occurs at the highly symmetric point X of its Brillouin zone (X is the center of the face of the cubic zone), and in this case the symmetry dictates $B_1=B_2$, $B_1'=B_2'=0$, and $a=0$ \cite{gork}.

\section{Magnetic susceptibility}

When $H$ is directed along the $p_1$ axis, the magnetic susceptibility $\chi$ (per unit volume) produced by the electron states in the vicinity of the  crossing point is described by the formula (\ref{1}) in which the constant factor $\Delta \chi$ has the form \cite{step}:
 \begin{equation}\label{15}
\Delta \chi=\frac{e^2}{12\pi^3\hbar c^2}
\frac{a'\beta}{B_1} f(\lambda),
 \end{equation}
where $\lambda =4 B_1 B_2 /\beta^2$,
 \begin{equation}\label{16}
f(\lambda)=(1-\lambda)\Delta \Phi -
\frac{\lambda}{2} \Delta \varphi ,
 \end{equation}
 \begin{eqnarray}\label{9}
\Delta\Phi\!\!&=&\!\!\int_{-\infty}^{\infty}\!\!d\tilde k_3 \int
\!\!\frac{dE\,{\rm sign}E}{(E^2-\tilde k_3^2)^{1/2}} \Big (
\frac{ 1-E}{E\,Y(1,E,\tilde k_3)} \nonumber \\ &+&\frac{Y(1,E,\tilde k_3)}{E^2 }\Big ), \\
\label{10}
\Delta\varphi\!\!&=&\!\!\int_{-\infty}^{\infty}\!\!d\tilde k_3 \int
\!\!\frac{dE\,{\rm sign}E}{(E^2-\tilde k_3^2)^{1/2}\,
 Y(1,E,\tilde k_3)} ,
 \end{eqnarray}
and
\begin{eqnarray}\label{11}
Y^2(1,E,\tilde k_3)\!\!&=&\!\!(1-E)^2 - \lambda
(E^2-\tilde k_3^2).
\end{eqnarray}
In formulas (\ref{9}) and (\ref{10}), the integration with respect to the dimensionless variable $E$ is carried out over the region defined by the conditions,
 \begin{eqnarray}\label{14}
 E^2-\tilde k_3^2 \ge 0, \\
 Y^2(1,E,\tilde k_3)\ge 0. \label{14a}
  \end{eqnarray}

\begin{figure}[tbp] 
 \centering  \vspace{+9 pt}
\includegraphics[scale=.50]{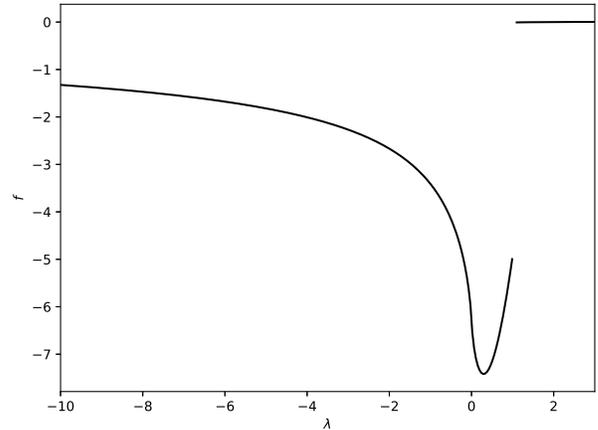}
\caption{\label{fig1} The function $f(\lambda)$ calculated with Eq.~(\ref{10a}).
 } \end{figure}

It was shown in Ref.~\cite{step} that $f(\lambda)=0$ at $\lambda>1$, and $f(0)=-2\pi$. We now calculate the function $f(\lambda)$ for an arbitrary value of $\lambda$. Let us define the range of the integration in formulas (\ref{9}) and (\ref{10}) only by condition  (\ref{14}). Then, $\Delta\Phi$ and $\Delta\varphi$ are the complex quantities, and condition (\ref{14a}) can be taken into account if we take the real part of these complex $\Delta\Phi$, $\Delta\varphi$.
Instead of the variable $\tilde k_3$, let us introduce the variable $\alpha$ in the above integrals:
 \[
 E^2-\tilde k_3^2=E^2\sin^2\!\!\alpha,
 \]
where $0\le \alpha \le \pi/2$. In this case, $Y^2=1-2E+c(\alpha)E^2$ with
  \[
  c(\alpha)\equiv 1-\lambda\sin^2\alpha.
  \]
The integration over $E$ is carried out explicitly. Ultimately, we arrive at
 \begin{eqnarray}\label{10a}
f(\lambda)&=&-2\pi(1-\lambda) -\lambda\, {\rm Re}\left[\int_{0}^{\pi/2}\!\!d\alpha \Lambda(\alpha) \right]\nonumber \\
&+& 2(1-\lambda){\rm Re}\left[\int_{0}^{\pi/2}\!\!d\alpha (c(\alpha)-1)\Lambda(\alpha) \right],
\end{eqnarray}
where
 \begin{eqnarray*}
\Lambda(\alpha)\equiv \frac{1}{\sqrt{c(\alpha)}}\ln \frac{\sqrt{c(\alpha)}+1}{\sqrt{c(\alpha)}-1}.
\end{eqnarray*}
Calculating the integrals in Eq.~(\ref{10a}) numerically, we obtain the function $f(\lambda)$ shown in Fig.~\ref{fig1}. Interestingly, this function exhibits the jump at $\lambda=1$. As was mentioned in Ref.~\cite{step}, the cases $\lambda < 1$ and $\lambda >1$ differ in the character of the transformation of the Fermi surface when the Fermi level $\zeta$ crosses the energy $\varepsilon_{\rm cr}$.
We now discuss the difference between the cases $\lambda < 1$ and $\lambda >1$ in more detail.

\begin{figure}[tbp] 
 \centering  \vspace{+9 pt}
\includegraphics[scale=0.8]{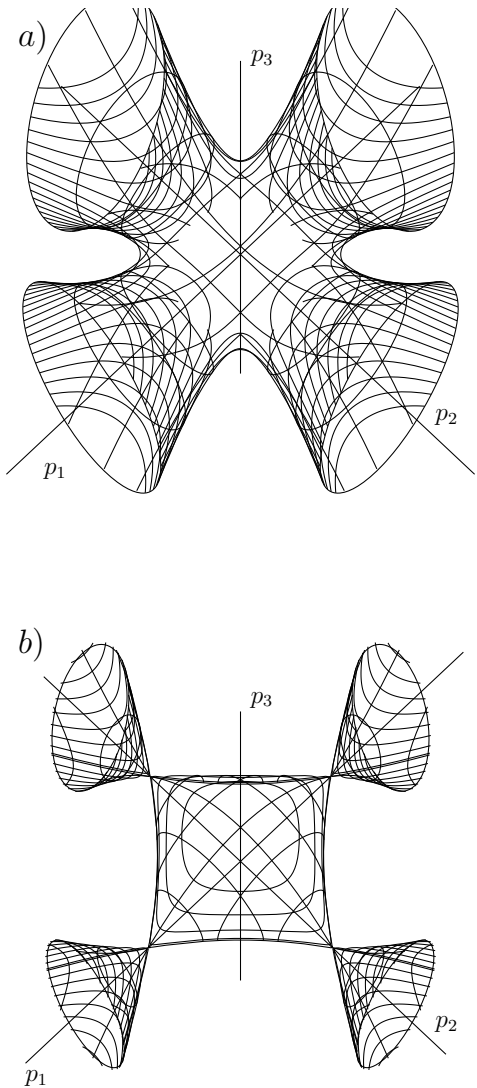}
\caption{\label{fig2} The Fermi surface at $0<\lambda<1$ for $\zeta-\varepsilon_{\rm cr}<0$ (a) and for $\zeta-\varepsilon_{\rm cr}>0$ (b). For definiteness, we assume that $B_1=B_2>0$ here. The central part of the self-intersecting Fermi surface in (b) corresponds to the electrons in the band ``$c$'', whereas its other parts, as well as the surface shown in (a), are produced by the holes of the band ``$v$''.
 } \end{figure}

\begin{figure}[tbp] 
 \centering  \vspace{+9 pt}
\includegraphics[scale=0.8]{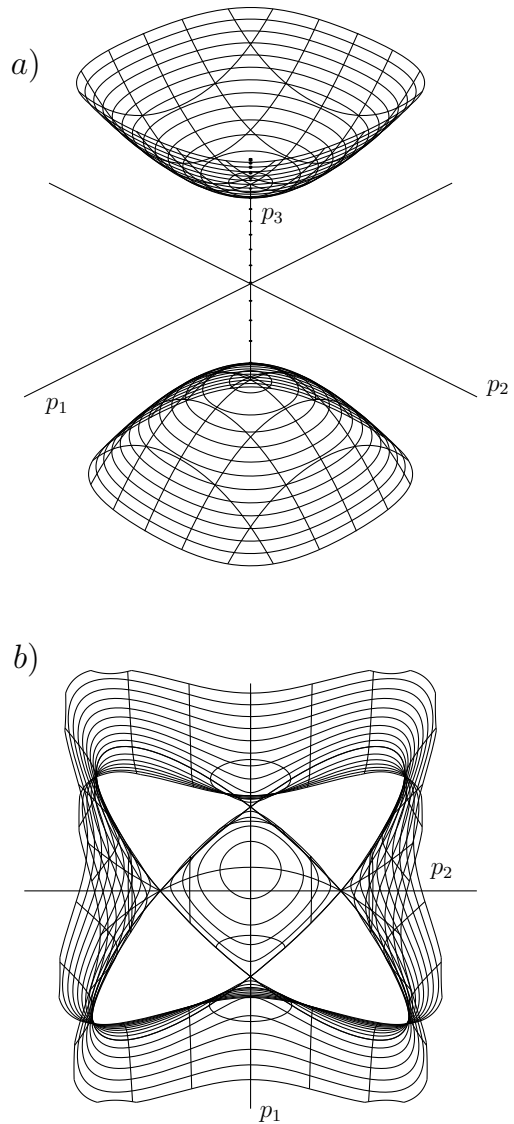}
\caption{\label{fig3}
The Fermi surface at $\lambda>1$ for $\zeta-\varepsilon_{\rm cr}<0$ (a) and for $\zeta-\varepsilon_{\rm cr}>0$ (b). As well as in Fig.~\ref{fig2},  $B_1=B_2>0$ here, and only the cental part of the self-intersecting Fermi surface shown in (b) corresponds to the electrons in the band ``$c$".
 } \end{figure}

At $\lambda<1$, the electron topological transition at $\zeta=\varepsilon_{\rm cr}$ is the break of the neck connecting the four tube-like surfaces of the band ``$v$'' and the concurrent appearance of the new pocket of the band ``$c$'' \cite{m-sh20} (Fig~\ref{fig2}). On the other hand, at $\lambda>1$, synchronously with the appearance of the new pocket of the band ``$c$'', the two separate surfaces of the band ``$v$'' merge at $\zeta= \varepsilon_{\rm cr}$, i.e., the neck of the joint surface is produced (Fig.~\ref{fig3}). In other words, for $\lambda < 1$ and $\lambda >1$, the two {\it different} topological transitions occur although the self-intersecting surfaces appear for both of them. This explains the jump in the function $f(\lambda)$.

To clarify the zero value of $f(\lambda)$ at $\lambda>1$, compare  dependences of the density of state on the Fermi level $\zeta$  for various electron topological transitions.
Let the Fermi surface of an electron energy band $i$ exhibit a transformation when the Fermi level $\zeta$ crosses the energy of the  transition, $\varepsilon_{c}$. Then, the density of electron states of this band, $\nu_i(\zeta)$,  can be represented in the form \cite{lif,lak,var,var1},
 \begin{equation}\label{nu}
 \nu_i(\zeta)=\nu_i^{\rm sm}(\zeta)+\nu_i^{\rm sp}(\zeta),
 \end{equation}
where $\nu_i^{\rm sm}(\zeta)$ is a smooth function of $\zeta$ in the vicinity of the energy $\varepsilon_{c}$, and
$\nu_i^{\rm sp}(\zeta)$ is the specific part of the density of states that differs from zero only above (or below) $\varepsilon_{c}$. In the case of the $2\frac{1}{2}$-order transition, one has $\nu_i^{\rm sp}(\zeta)\propto |\zeta-\varepsilon_c|^{1/2}$ when a new pocket of the Fermi surface appears or its neck breaks. In the case of the
$3\frac{1}{2}$-order transition, the two bands ``$c$'' and ``$v$'' contacting along the line have the common transition energy $\varepsilon_c$, and $\nu_i$ for each of  these two bands has the form of Eq.~(\ref{nu}) with $\nu_c^{\rm sp}(\zeta)=\nu_v^{\rm sp}(\zeta) \propto |\zeta- \varepsilon_c|^{3/2}$ \cite{m-sh14,m-sh-jltp}. These $\nu_c^{\rm sp}$, $\nu_v^{\rm sp}$ differ from zero  when the self-intersecting Fermi surface appears. For the transition at the crossing point of the two band-contact lines, one has $\varepsilon_c= \varepsilon_{\rm cr}$, and at $\lambda<1$, the total specific part of the density of state for the bands ``$c$'' and ``$v$'', $\nu_c^{\rm sp}(\zeta)+\nu_v^{\rm sp}(\zeta)$, differs from zero when the self-intersection of the surface occurs. On the other hand, at $\lambda>1$, we obtain that  $\nu_c^{\rm sp}(\zeta)=-\nu_v^{\rm sp}(\zeta)$, and the total density of states for the two bands, $\nu_c(\zeta)+\nu_v(\zeta)$, is a smooth function above and below the energy $\varepsilon_c=\varepsilon_{\rm cr}$ (with the infinite derivative at this energy). In other words, although  the topological transitions occur for each of the bands at $\lambda>1$, the appropriate specific parts of $\nu$ compensate each other. In spite of the fact that the orbital magnetic susceptibility $\chi$ is not proportional to the density of state, our results show that the zero value of $\Delta\chi$ at $\lambda>1$ can be interpreted as a similar compensation of the appropriate contributions to the susceptibility. Formula (\ref{1}) also demonstrate that the magnetic susceptibility can be the effective probe of the topological transition occurring at $\zeta=\varepsilon_{\rm cr}$  since $\Delta\chi$ is not small as compared to $\chi_0$, and this $\Delta\chi$ is independent of the temperature.

\section{Susceptibility of ${\rm V}_3{\rm Si}$}

\begin{figure}[tbp] 
 \centering  \vspace{+9 pt}
\includegraphics[scale=1.]{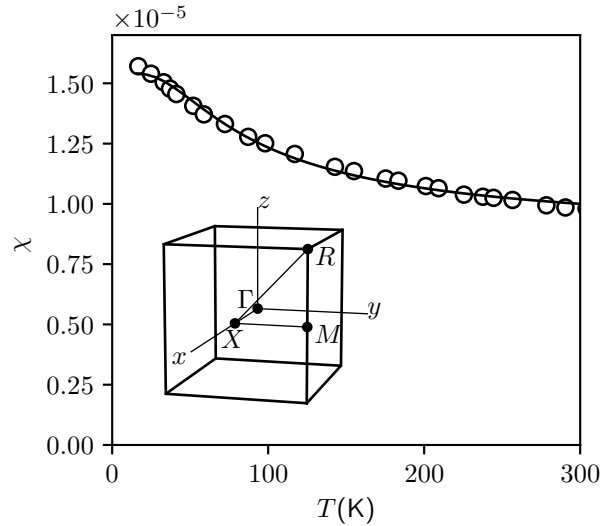}
\caption{\label{fig4} The temperature dependence of magnetic susceptibility $\chi$ of V$_3$Si according to formulas (\ref{1}) and (\ref{15a}) (the solid line). Here $\chi_0\approx 1.54\cdot 10^{-5}$, $\varepsilon_{\rm cr}-\zeta \approx 122$ K, $\Delta \chi\approx -1.36\cdot 10^{-5}$; at $a'=2\cdot10^5$ m/s this value of $\Delta \chi$ corresponds to $\lambda\approx 0.0024$. The cycles mark the experimental data \cite{maita} which have been recalculated per unit volume of the sample. Inset: The Brillouin zone of V$_3$Si and its characteristic points.
 } \end{figure}

The magnetic susceptibility $\chi$ of V$_3$Si above the temperature of its superconducting transition  was measured many years ago, and essential temperature dependence of $\chi$ was discovered  \cite{clog,maita}; see Fig.~\ref{fig4}. Gorkov \cite{gork} suggested that the chemical potential in V$_3$Si lies near a degeneracy energy of two bands at the X point of the Brillouin zone of this cubic crystal, and these bands are described by formulas (\ref{2}) and (\ref{3}) (with $B_1=B_2$, $B_1'=B_2'=a=0$). He obtained that $\chi$ should logarithmically depends on temperature $T$ in this case. However, this explanation of the temperature dependence of $\chi$ meets with difficulties since the logarithmic term in the susceptibility is expected to be small \cite{gork,vonsovsk}. Moreover, the calculation of the magnetic susceptibility in Ref.~\cite{gork} was not complete because only the linear terms with respect to $H$ were taken into account in the Hamiltonian of  electrons in the magnetic field. In the calculation of the magnetic susceptibility, $\chi=-(1/H)\partial \Omega/\partial H$, the electron thermodynamic potential $\Omega$ is found in the second order in $H$. Hence, for the correct calculation of $\chi$, it is necessary to take into account both the linear and quadratic terms in the electron Hamiltonian. In this case, the calculation of $\chi$ leads to formula (\ref{1}) since the point X is the crossing point of the two band-contact lines (the directions of these lines coincide with X-R lines in the Brillouin zone of V$_3$Si, Fig.~\ref{fig4}).

For the crossing point with $B_1=B_2\equiv B$ and $B_1'=B_2'=a=0$,
the factor $\Delta\chi$ in formula (\ref{1}) does not depend on the direction of the magnetic field in the $1-2$ plane. On the other hand, the magnetic susceptibility is practically independent of $\zeta$ and $T$ when $H$ is along the $3$ axis \cite{step}. Taking into account that there are six X points in the Brillouin zone of   V$_3$Si, we find that the magnetic susceptibility of this cubic crystal, as expected, does not depend on the direction of $H$ and is described by formula (\ref{1}) with the following $\Delta \chi$:
 \begin{equation}\label{15a}
\Delta \chi=\frac{2e^2}{3\pi^3\hbar c^2}
\frac{a'}{\sqrt{\lambda}} f(\lambda),
 \end{equation}
where we have taken into account that $\lambda=4B^2/\beta^2$ now. Since this $\lambda$ is positive, the factor $\Delta \chi$ is nonzero and negative only if $\lambda<1$, Fig.~\ref{fig1}.
Moreover, we shall see below that $\lambda$ seems small in V$_3$Si, i.e.,   $f(\lambda)\sim -2\pi$ in Eq.~(\ref{15a}).

In the experiments \cite{clog,maita}, the magnetic susceptibility decreases with increasing $T$. Then, according to formula (\ref{1}),  we should assume that $\varepsilon_{\rm cr}-\zeta >0$, i.e., the chemical potential lies below $\varepsilon_{\rm cr}$. The parameter $a'$ can be estimated from the band-structure calculations along the $\Gamma$-X axis of V$_3$Si. Taking $0.472$ nm as the linear size of the cubic cell of this crystal \cite{testardi}, we obtain the crude estimate, $a'\approx 2\cdot10^5$ m/s, from Fig.~4 of Ref.~\cite{matt}. Then, $\Delta \chi\approx 1.05\cdot 10^{-7}f(\lambda)/\sqrt{\lambda}$, and the parameter $\lambda$ can be found from a fit of dependence (\ref{1}) to the appropriate experimental data.

Figure \ref{fig4} shows our fit of the magnetic susceptibility calculated with formulas (\ref{1}), (\ref{15a}) to the experimental data of Ref.~\cite{maita}. (These data have been recalculated per unit volume, using the following values for V$_3$Si: $1$ mole $=180.91$ g and the density $\rho=5.7$ g/cm$^3$). This fit gives $\chi_0\approx 1.54\cdot 10^{-5}$, $\varepsilon_{\rm cr}-\zeta \approx 122$ K and $\Delta \chi\approx -1.36\cdot 10^{-5}$. The obtained value of $\Delta \chi$ corresponds to $\lambda\approx 2.4\cdot 10^{-3}$ if $a'\approx 2\cdot 10^5$ m/s. Note that the value of $\varepsilon_{\rm cr}-\zeta$ is small as compared to that found in the band structure calculations ($\sim 24 $ mRy) \cite{matt75,matt}. However, the error of these calculation is just of the order of $20$ mRy \cite{matt75}.

\begin{figure}[tbp] 
 \centering  \vspace{+9 pt}
\includegraphics[scale=1.]{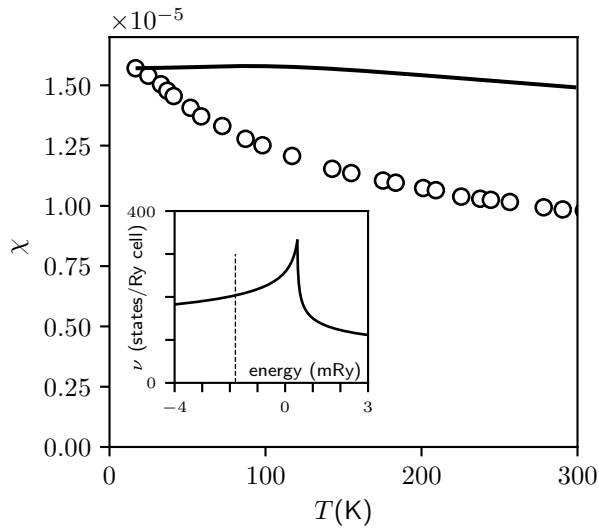}
\caption{\label{fig5} The function $\chi(T)+\bar{\chi}$ where $\bar{\chi}=8.05\cdot 10^{-6}$ and $\chi(T)$ is described by Eqs.~(\ref{13b}), (\ref{14b}) with $\zeta=-1.8$ mRy (the solid line). The cycles mark the experimental data \cite{maita} shown in Fig.~\ref{fig4}. Inset: The dependence of the density of states $\nu$ on the energy \cite{matt}, Eq.~(\ref{13b}),  in the vicinity of the Fermi level $\zeta$ that is indicated by the vertical dashed line.
 } \end{figure}   

Let us discuss the possibility of a large value of $|\zeta-\varepsilon_{\rm cr}|$ in more detail. Since the contribution of the electron states near the points X to the susceptibility cannot explain its noticeable temperature dependence if $\varepsilon_{\rm cr}$ is far from the Fermi level, we now consider another possible explanation of this dependence. This explanation was suggested many years ago \cite{clog61,labbe}, and it is based on existence of a peak in the dependence of the density of electron states $\nu$ on the energy $\varepsilon$ near the Fermi level $\zeta$. According to Fig.~12a of Ref.~\cite{matt}, this peak really exists. In the vicinity of the peak, we find that the density of states is well approximated by the function (Fig.~\ref{fig5}),
 \begin{eqnarray}\label{13b}
\nu(\varepsilon)=\frac{235.08}{(0.57-\varepsilon)^{1/6}}, \ \ \varepsilon\le 0.439, \nonumber \\
\nu(\varepsilon)=\frac{134.45}{(\varepsilon-0.428)^{1/5}}, \ \ \varepsilon\ge 0.439,
\end{eqnarray}
where $\varepsilon$ is measured in mRy, and $\nu$ in states per Ry and the crystal cell. The zero value of $\varepsilon$ is chosen in Ref.~\cite{matt} so that $\zeta=-1.8$ mRy. Neglecting a small shift of $\zeta$ with the temperature $T$, we obtain the following expression for the spin magnetic susceptibility produced by the peak in the density of states:
 \begin{eqnarray}\label{14b}
\chi(T)=\mu_B^2\int\frac{\nu(\varepsilon)d\varepsilon }{4T\cosh^2[(\varepsilon-\zeta)/2]},
\end{eqnarray}
where $\mu_B$ is the Bohr magneton. In Fig.~\ref{fig5} we show $\chi(T)$ calculated according to Eqs.~(\ref{13b}) and (\ref{14b}), with the constant term $\bar{\chi}=8.05\cdot 10^{-6}$ being added to $\chi(T)$ for the calculated magnetic susceptibility to coincide with experimental data at low temperatures. This additional term may be ascribed to an orbital magnetic susceptibility associated with filled electron bands. It is clear from Fig.~\ref{fig5} that the peak in density of states cannot explain the experimental temperature dependence of the magnetic susceptibility. This result supports our assumption that the temperature dependence of the magnetic susceptibility of V$_3$Si is due to the crossing of the band-contact lines in the vicinity of the Fermi level. The spin susceptibility described by Eq.~(\ref{14b}) can be considered as a part of the background term $\chi_0$.

\end{document}